\documentclass[aps,prd,reprint,superscriptaddress,nofootinbib,showkeys]{revtex4-2}
\usepackage[T1]{fontenc}
\usepackage[utf8]{inputenc}
\usepackage{lmodern}
\usepackage{amsmath,amssymb}
\usepackage{graphicx}
\usepackage{xcolor}
\usepackage{pgfplots}
\usepackage{microtype}
\usepackage[unicode,hidelinks]{hyperref}
\usepackage{url}
\usepackage{booktabs}

\begin{document}

\title{Quantifying the TKM-IKT ambiguity: translating the theoretical benchmark into experimental discriminants for future colliders}

\author{Zardo Becker, G.}
\affiliation{Physics Institute, Federal University of Santa Catarina (UFSC), Florianópolis/SC, Brazil}
\affiliation{Physics Institute, The Ohio State University (OSU), Columbus/OHIO, United States}
\email{zardobecker@gmail.com}
\author{Benno Bodmann}
\affiliation{International Center for Relativistic Astrophysics Network headquartered on Universidade Federal do Rio Grande do Sul (ICRANet/UFRGS) , Porto Alegre, Brazil}
\email{benno.bodmann@gmx.de}
\author{C\'esar A. Zen Vasconcellos}
\email{cesaraugustozenvasconcellos@gmail.com}
\affiliation{International Center for Relativistic Astrophysics Network (ICRANet), Pescara, Italy}
\begin{abstract}
{This paper, the second of two companion works, translates the theoretical benchmark established in \citet{ZardoBecker2026a} into quantitative predictions for high-energy $\gamma\gamma$ interactions. We simultaneously confront the TKM \citep{tkm_2002} and IKT \citep{Iancu2004} dipole cross-section prescriptions across five representative saturation models (GBW \citep{GBW_1998}, IIMS \citep{IIMS_2007}, rcBK \citep{RCBK2011}, AGBS \citep{Amaral_2007a}, WYKWC \citep{WYKWC_2021:first_eq}). Evaluating real and virtual cross sections, $F_2^\gamma$, and heavy-quark production, we demonstrate that the TKM--IKT discrepancy is an irreducible structural ambiguity reaching $48\pm12\%$ at LEP energies and expanding into the TeV regime. We propose three experimental discriminants for future colliders (ILC, CLIC, FCC-ee)—emphasizing exclusive $\gamma\gamma \to J/\psi + J/\psi$ production where quadratic amplitude scaling ($[\mathcal{N}]^2$) magnifies the divergence—elevating these comparisons from parameter choices to rigorous falsification tests of the foundational premises. Finally, an analytical universality test using GBW DIS parameters establishes the quantitative baseline required to falsify the underlying dipole--target analogy.}
\end{abstract}


\maketitle
\tableofcontents

\section{Introduction}

In the companion paper~\citep{ZardoBecker2026a}, we established the theoretical basis for two-photon interactions at high energies and, more importantly, we defined a rigorous theoretical benchmark by systematically mapping the conceptual limitations of the current state of the art: the heuristic nature of the TKM~\citep{tkm_2002} and IKT~\citep{Iancu2004} prescriptions for the dipole–dipole cross section, their violation of exchange symmetry, the unproven assumption of universality of the dipole–proton amplitude $\mathcal{N}(r,Y)$, and the approximations inherent in the homogeneous-disk treatment. That work delineated the \textit{envelope of theoretical uncertainty} within which any phenomenological description must lie. These limitations were discussed in detail and categorised in the companion paper~\citep{ZardoBecker2026a}, where they were framed as the defining features of the theoretical benchmark. Moving beyond a purely instrumental approach, the present paper assumes this benchmark to evaluate how these theoretical constraints manifest in physical observables, providing a critical baseline to confront upcoming experimental data.

In view of this theoretical benchmark, the central questions addressed in this paper unfold along two fronts: first, to rigorously quantify the magnitude of the phenomenological spread generated by the heuristic TKM~\citep{tkm_2002} and IKT~\citep{Iancu2004} prescriptions when applied to physical observables; and second, to investigate whether future experiments will be capable of going beyond a mere parameter choice, testing whether they can falsify the foundational premises of the benchmark itself. We argue that the ambiguity between TKM and IKT does not constitute a mere numerical noise susceptible to being eliminated by artificial laboratory tunings; rather, it directly reflects the structural exhaustion of approaches based on static geometries and homogeneous disks. Consequently, the discriminants we propose assume the character of rigorous falsification tests: should next-generation data from linear and circular colliders lie outside the envelope bounded by the prescriptions and saturation models, this will evidence that the classical hypotheses of universality, exchange symmetry, and the neglect of the impact parameter have reached their breaking point, demanding a profound reformulation in the description of the photon's parton density.

Specifically, if future experimental data fall outside the TKM--IKT predictive envelope, this will indicate that at least one of the fundamental premises systematically mapped in the companion work~\citep{ZardoBecker2026a} has been violated: (i) the hypothesis of universality of the dipole amplitude extracted from DIS processes; (ii) the validity of the asymmetric evolution restricted to only a single dipole in rapidity, which would require a fully symmetric treatment depending on $Y_1 + Y_2$; or (iii) the homogeneous-disk approximation, which disregards impact-parameter dependence. In any of these scenarios, a deep conceptual revision of the analogy between dipole and target will become unavoidable, paving the way beyond traditional heuristic prescriptions.

To translate this theoretical benchmark into a quantitative phenomenological baseline, we compute the primary observables for two-photon collisions—including the total real and virtual cross sections $\sigma^{\gamma\gamma}$ and $\sigma^{\gamma^*\gamma^*}$, the photon structure function $F_2^\gamma$, and inclusive heavy-quark production ($c\bar{c}, b\bar{b}$)—employing all five saturation models [GBW~\citep{GBW_1998}, IIMS~\citep{IIMS_2007}, rcBK~\citep{RCBK2011}, AGBS~\citep{Amaral_2007a}, and WYKWC~\citep{WYKWC_2021:first_eq}] alongside both dipole cross-section prescriptions [TKM~\citep{tkm_2002} and IKT~\citep{Iancu2004}]. Confronting these predictions with historical LEP data reveals that the systematic spread between TKM and IKT is by no means an artifact of parameter tuning, but rather a genuine structural ambiguity that reaches $48\pm12\%$ at $W_{\gamma\gamma} = 200\text{ GeV}$ and expands systematically into the multi-TeV domain. Furthermore, using the original DIS parameters for the GBW model~\citep{GBW_1998} without re-adjustment, we find that the resulting cross sections are smaller by a factor $\sigma_0^{\text{DIS}}/\sigma_0^{\gamma\gamma} \approx 0.84$; given the typical experimental uncertainties of 15–20\% at LEP, this keeps the predictions within a factor of 1.5–2 of the data, demonstrating that the universality of the dipole amplitude is approximately satisfied while leaving a minor residual discrepancy for future precision measurements.

Heavy-quark production via fragmentation is estimated using a simple $K$-factor method based on $\gamma p$ data from the OPAL and L3 collaborations. We adopt $K_c = 1.5$ for charm and $K_b = 1.2$ for bottom, resulting in bands that indicate fragmentation may enhance the direct cross section by 20–50\% for charm and 20–30\% for bottom. These are order-of-magnitude estimates; a full implementation of fragmentation functions in the dipole formalism is left for future work, but our estimates serve as a conservative indicator of the potential importance of these contributions.

To discriminate between the two physical pictures embodied by TKM (a dense+small photon) and IKT (a dilute+large photon), we propose three clear experimental observables that future $e^+e^-$ colliders with $\sqrt{s}\gtrsim 1$ TeV can measure: (i) the energy growth of the total cross section $\sigma^{\gamma\gamma}$, where TKM predicts a factor 2–3 larger value than IKT at $W=1$ TeV; (ii) the charm-to-bottom production ratio $R_{c/b} = \sigma^{c\bar{c}}/\sigma^{b\bar{b}}$, which is approximately 30\% higher in TKM due to its emphasis on small dipoles; and (iii) exclusive vector meson pair production $\gamma\gamma \to J/\psi J/\psi$, which scales as $[\mathcal{N}(r,Y)]^2$ and thus amplifies the differences – TKM yields a cross section an order of magnitude larger than IKT at high $W$. These discriminants are not merely tools for choosing between models; they are tests of the benchmark itself. A measurement outside the TKM–IKT envelope would signal a violation of universality, the insufficiency of the asymmetric evolution treatment, or the need for a photon-specific saturation scale.

The numerical stability of all calculations has been verified systematically by varying the integration tolerances, the angular discretisation, and the lower cutoff $r_{\min}$ for a representative point. All observed variations are at least two orders of magnitude smaller than the TKM–IKT spread, ensuring that the conclusions are robust and independent of numerical artifacts.

We emphasise that the TKM~\citep{tkm_2002} and IKT~\citep{Iancu2004} prescriptions remain phenomenological. Their heuristic nature, the breaking of exchange symmetry when the two dipoles have comparable sizes, and the untested assumption of universality of the dipole–proton amplitude $\mathcal{N}(r,Y)$ were systematically mapped in the companion paper~\citep{ZardoBecker2026a}. These limitations are not treated here as defects to be corrected within the same framework; rather, they define the theoretical envelope — the benchmark — against which future data will be compared. A genuine first-principles derivation of $\sigma_{dd}$ from the JIMWLK functional is still lacking, and the asymmetric evolution used here (evolving only one dipole) should eventually be replaced by a symmetric treatment depending on $Y_1+Y_2$. The homogeneous-disk approximation (neglecting impact-parameter dependence) may also be insufficient for exclusive processes. The discriminants proposed in this paper — $\sigma^{\gamma\gamma}$, $R_{c/b}$, and exclusive $J/\psi J/\psi$ production — are explicitly designed to test these very assumptions. If future data fall outside the TKM–IKT envelope, it will indicate that at least one of these heuristic assumptions is violated, prompting a revision of the dipole–target analogy. These open questions define a clear roadmap for future theoretical work, beyond the present phenomenological benchmark.

This paper is organized as follows. Sec.~\ref{sec:results} presents the results for the real and virtual photon cross sections, the photon structure function, heavy-quark production, the flavour decomposition, and the non-integrated analysis. Sec.~\ref{sec:syst} quantifies the systematic uncertainties and discusses the experimental discriminants in detail. Finally, Sec.~\ref{sec:conclusions} presents our conclusions and outlines the open questions that define the roadmap for future work. All technical details of the dipole formalism, the TKM/IKT prescriptions, and the saturation models can be found in the companion paper~\citep{ZardoBecker2026a}.

\section{Results}\label{sec:results}

Here we present the results for: (i) the real cross section $\sigma^{\gamma \gamma}$, where both photons have zero virtuality ($Q^2_{1,2} = 0$); (ii) the inclusive heavy quark photoproduction processes $\gamma \gamma \rightarrow c\bar{c}X$ and $\gamma \gamma \rightarrow b\bar{b}X$ from real photons; (iii) the analysis of quark flavor contributions compared with low-energy components (Reggeons and QPM); (iv) the virtual cross section $\sigma^{\gamma^\ast \gamma^\ast}$, where both photons have the same virtuality ($Q^2_1 = Q^2_2 = Q^2$); (v) the photon structure function $F_2^\gamma$, where one photon is real and the other is virtual; and (vi) the analysis of how the dipole--dipole cross section prescription relates dipole size dependence to observables involving dilute and dense partonic regimes in the photon.

The parameters of the dipole-proton amplitudes (the five models) were fitted to LEP data for $\sigma^{\gamma\gamma}$ (TT polarization) as described in the companion paper. The best-fit parameters for the complete contribution (VDM+QPM+gluonic) are listed in Table~\ref{tab:parameters_complete}, and for the pure gluonic contribution (using only the last four LEP points) in Table~\ref{tab:parameter_gluonic}. The mass of light quarks in the Reggeon contribution is $m_q=0.210$ GeV.

\begin{table}[t]%
\centering
\caption{Selected parameters of IKT and TKM for the complete contribution.}
\tabcolsep=0pt%
\begin{tabular*}{20pc}{@{\extracolsep\fill}lcccc@{\extracolsep\fill}}
\toprule
{Amplitude} &\multicolumn{2}{@{}c@{}}{{TKM}} & \multicolumn{2}{@{}c@{}}{{IKT}} \\\cmidrule{2-3}\cmidrule{4-5}
     {Model}  &   { $\Lambda_c$ [MeV]} & $\chi^2 /N$ & $m_l$ [Mev] & $\chi^2 /N$ \\
     \midrule
    AGBS & 176 & 1.337 & 205 & 1.065 \\
    Iancu2004 & 242 & 1.086 & 205 & 1.081 \\
    WYKWC & 168 & 0.967 & 210 & 1.114 \\
    GBW & 249 & 0.918 & 211 & 0.984 \\
    AAMQS & 216 & 0.906 & 205 & 1.019\\
\bottomrule
\end{tabular*}
\label{tab:parameters_complete}
\end{table}

\begin{table}[htbp]
\centering
\caption{Selected parameters for the gluonic contribution only.}
\tabcolsep=0pt%
\begin{tabular*}{20pc}{@{\extracolsep\fill}lcc@{\extracolsep\fill}}
\toprule
    {Amplitude Model} & {IKT}, {$\Lambda_c$ [MeV]} & {TKM}, {$m_l$ [Mev]}  \\
       \midrule
    AGBS & 160 & 185    \\
    Iancu2004 & 227 & 185     \\
    WYKWC & 150 & 185  \\
    GBW & 230 & 192  \\
    AAMQS & 218 & 182  \\
\bottomrule
\end{tabular*}
\label{tab:parameter_gluonic}
\end{table}

\subsection{Real photon cross section}
\begin{figure}[htbp]
\centering
\includegraphics[width=0.5\textwidth]{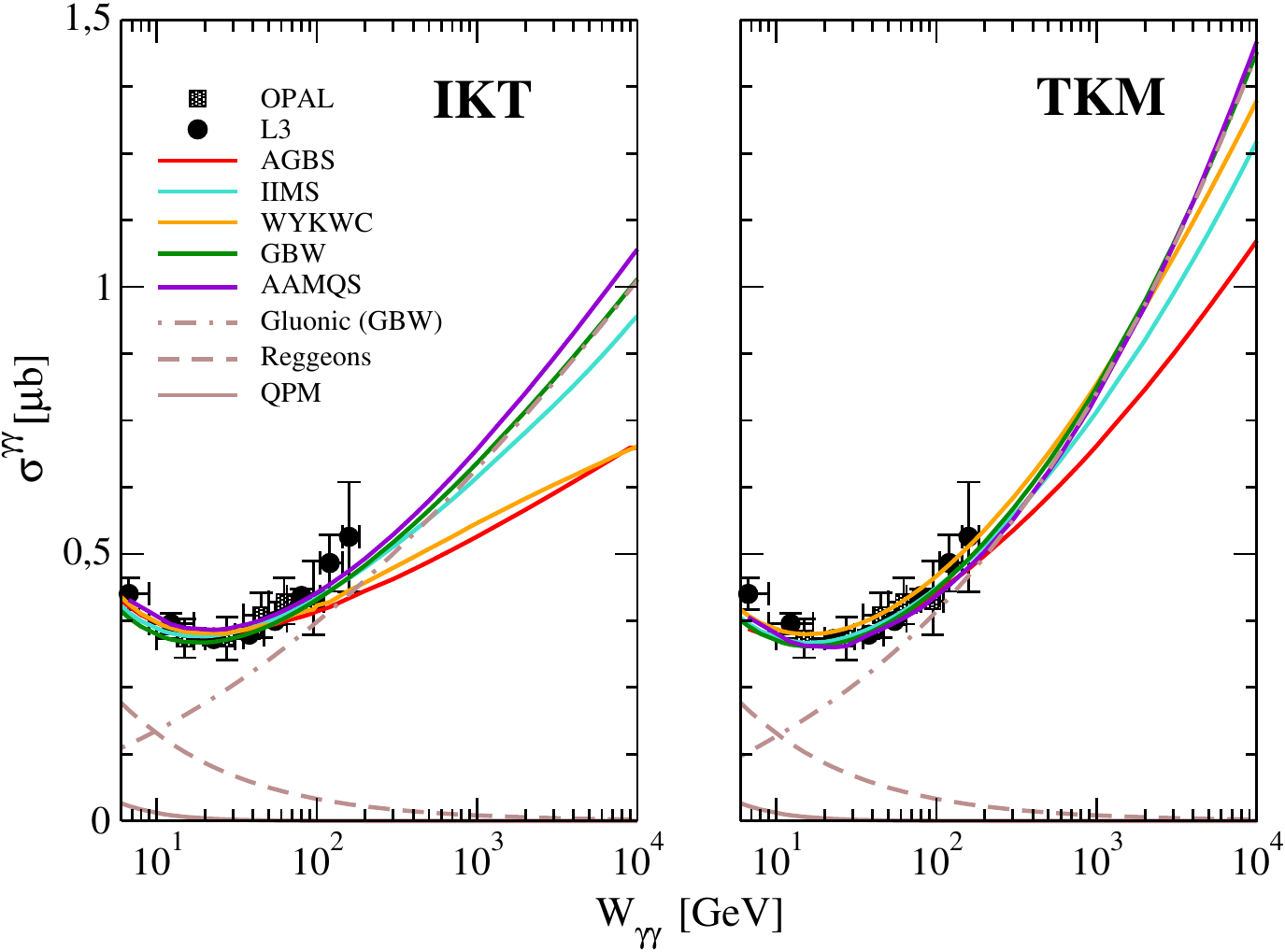}
\caption{Real photon cross section (complete) compared with LEP data. Left: IKT; right: TKM.}
\label{fig:SR_complete}
\end{figure}

Figure~\ref{fig:SR_complete} shows $\sigma^{\gamma\gamma}$ (including VDM, QPM and gluonic components) as a function of $W_{\gamma\gamma}$ for the IKT (left) and TKM (right) prescriptions. All models describe LEP data up to about 150 GeV. Above 200 GeV the curves separate: IKT shows a clear split between coordinate-space models (higher production) and momentum-space models (lower), while TKM gives higher and more grouped curves.

\begin{figure}[htbp]
\centering
\includegraphics[width=0.5\textwidth]{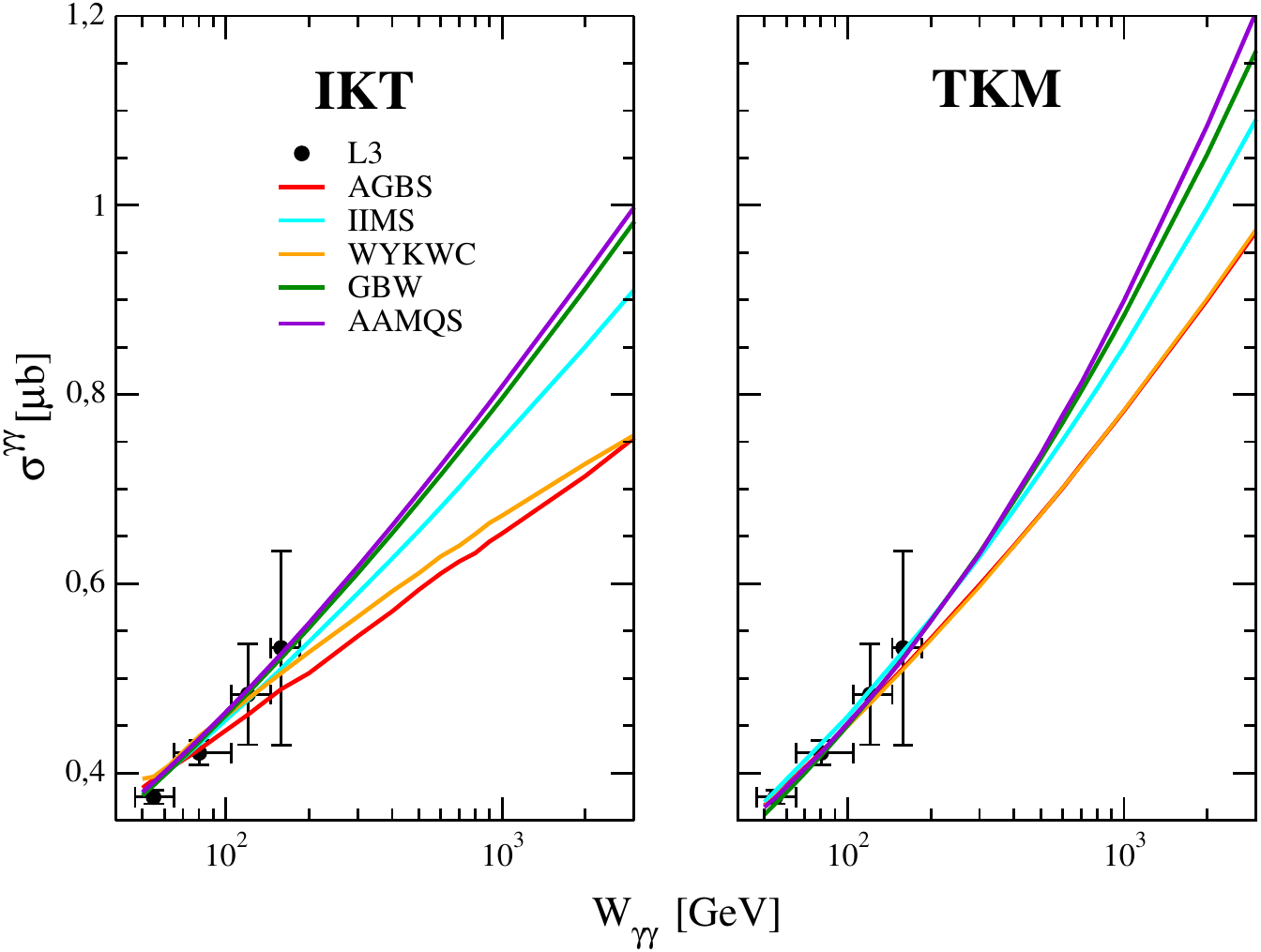}
\caption{Gluonic contribution only. Left: IKT; right: TKM.}
\label{fig:SR_gluonic}
\end{figure}

Figure~\ref{fig:SR_gluonic} shows the pure gluonic contribution. The parameters are re-fitted using the last four LEP points ($W>50$ GeV). The TKM prescription again yields a larger cross section at high energies.

\subsection{Test of universality}\label{testuniversality}

We test the hypothesis that the same dipole–proton amplitude $\mathcal{N}(r,Y)$, originally extracted from deep inelastic scattering (DIS) off protons, can be applied to $\gamma\gamma$ collisions without re‑adjustment. Using the original DIS parameters for the GBW model ($\sigma_0^{\text{DIS}}=23.03$ mb, $\lambda^{\text{DIS}}=0.288$, $x_0^{\text{DIS}}=3.04\times10^{-4}$) and keeping the same $m_l$ and $\Lambda_c$ as in the $\gamma\gamma$ fits, we compute $\sigma^{\gamma\gamma}$ and compare it with the re‑fitted predictions (Table~\ref{tab:parameter_gluonic}). The DIS‑tuned cross sections are smaller by a factor $\sigma_0^{\text{DIS}}/\sigma_0^{\gamma\gamma} \approx 0.84$. Given the typical experimental uncertainties of 15–20\% at LEP energies, this reduction still keeps the predictions within a factor 1.5–2 of the data.

We therefore conclude that, for the GBW model, the universality of the dipole amplitude is approximately satisfied. A full test of universality across all five saturation models (Iancu2004 \citep{Iancu2004}, rcBK \citep{RCBK2009, RCBK2011}, AGBS \citep{Amaral_2007a, Amaral_2021}, and WYKWC \citep{WYKWC_2021:first_eq, WYKWC_2022:Meson_production}) is left for future work, as the parametrizations of these models are more complex and their DIS parameters are less uniquely defined. For the purposes of this benchmark, the GBW result is taken as indicative of the approximate validity of the universality assumption. A mild re‑adjustment of $\Lambda_c$ or $m_l$ (as in Tables~\ref{tab:parameters_complete} and~\ref{tab:parameter_gluonic}) improves the quantitative agreement, but the residual discrepancy indicates that future precision data could help resolve the validity of universality across different models.

\subsection{Heavy-quark inclusive production}
\begin{figure}[htbp]
\centering
\includegraphics[width=0.46\textwidth]{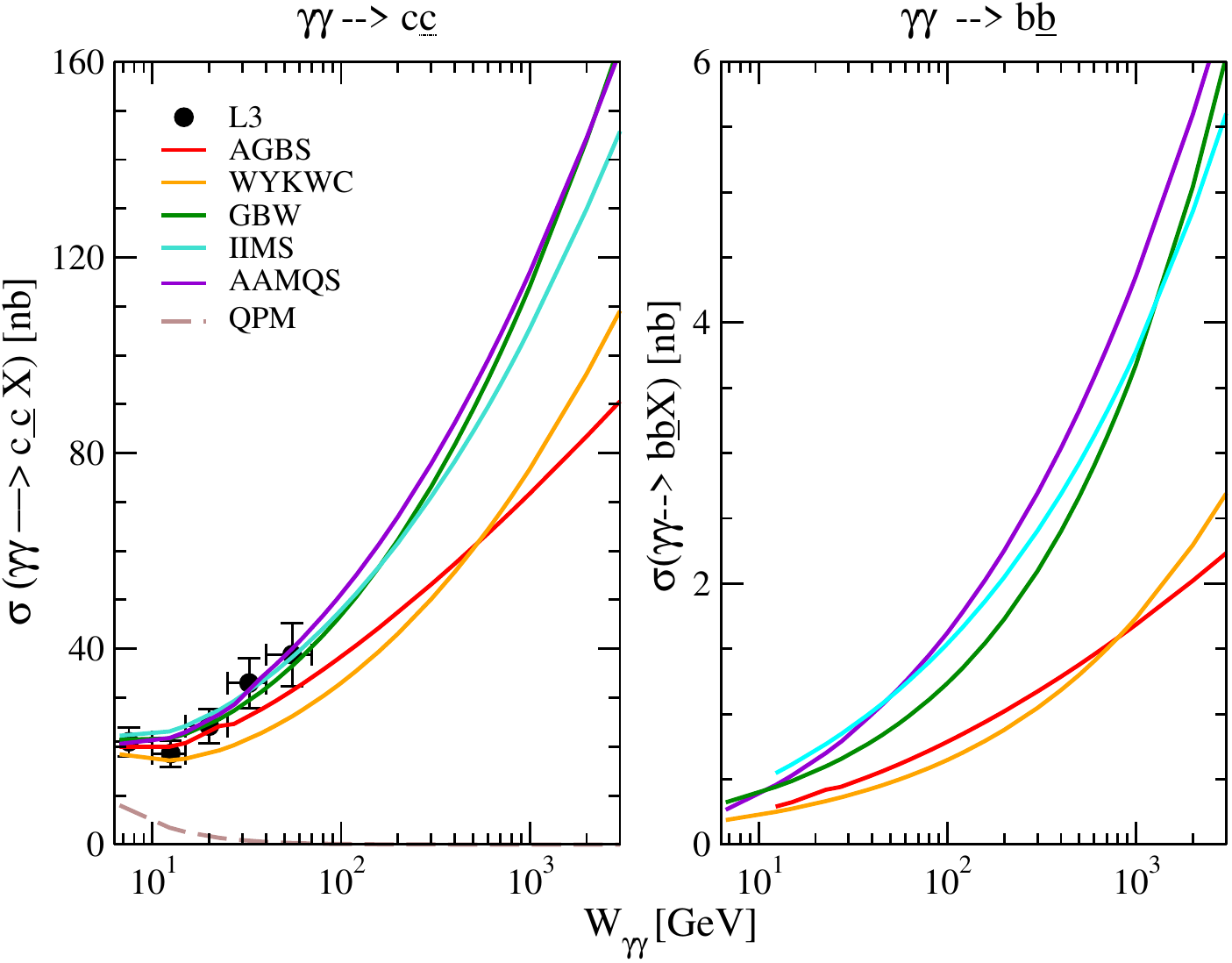}
\caption{Heavy-quark inclusive production. Solid lines: direct production computed within the dipole formalism; shaded bands: indicative estimates of fragmentation contributions based on \(K_c = 1.5\) and \(K_b = 1.2\), taken from \(\gamma p\) data analyses (OPAL, L3). These bands are order-of-magnitude estimates and should be interpreted as indicative, not as precise predictions.}
\label{fig:heavy_quarks}
\end{figure}

Figure~\ref{fig:heavy_quarks} shows the cross sections for $\gamma\gamma\to c\bar{c}X$ (left) and $\gamma\gamma\to b\bar{b}X$ (right) computed with the IKT prescription. The solid lines are the direct production (lower bound). The shaded bands indicate an estimated fragmentation contribution using $K_c=1.5$ for charm and $K_b=1.2$ for bottom, based on $\gamma p$ data. The fragmentation enhancement may be 20-50\% for charm and 20-30\% for bottom.

Heavy-quark production via fragmentation is estimated using a simple \(K\)-factor method. The values \(K_c = 1.5\) for charm and \(K_b = 1.2\) for bottom are taken from existing analyses of \(\gamma p\) data (OPAL and L3 collaborations), where such factors were introduced to account for the enhancement of heavy-quark cross sections due to fragmentation. These values are not derived from the dipole formalism used in this work; they are included here as order-of-magnitude estimates to illustrate the potential importance of fragmentation contributions. The resulting bands in Fig.~\ref{fig:heavy_quarks} (20–50\% for charm, 20–30\% for bottom) reflect the typical spread found in those \(\gamma p\) analyses. A full implementation of fragmentation functions within the dipole formalism, which would be required for precision predictions, is left for future work.

\begin{figure}[htbp]
\centering
\includegraphics[width=0.4\textwidth]{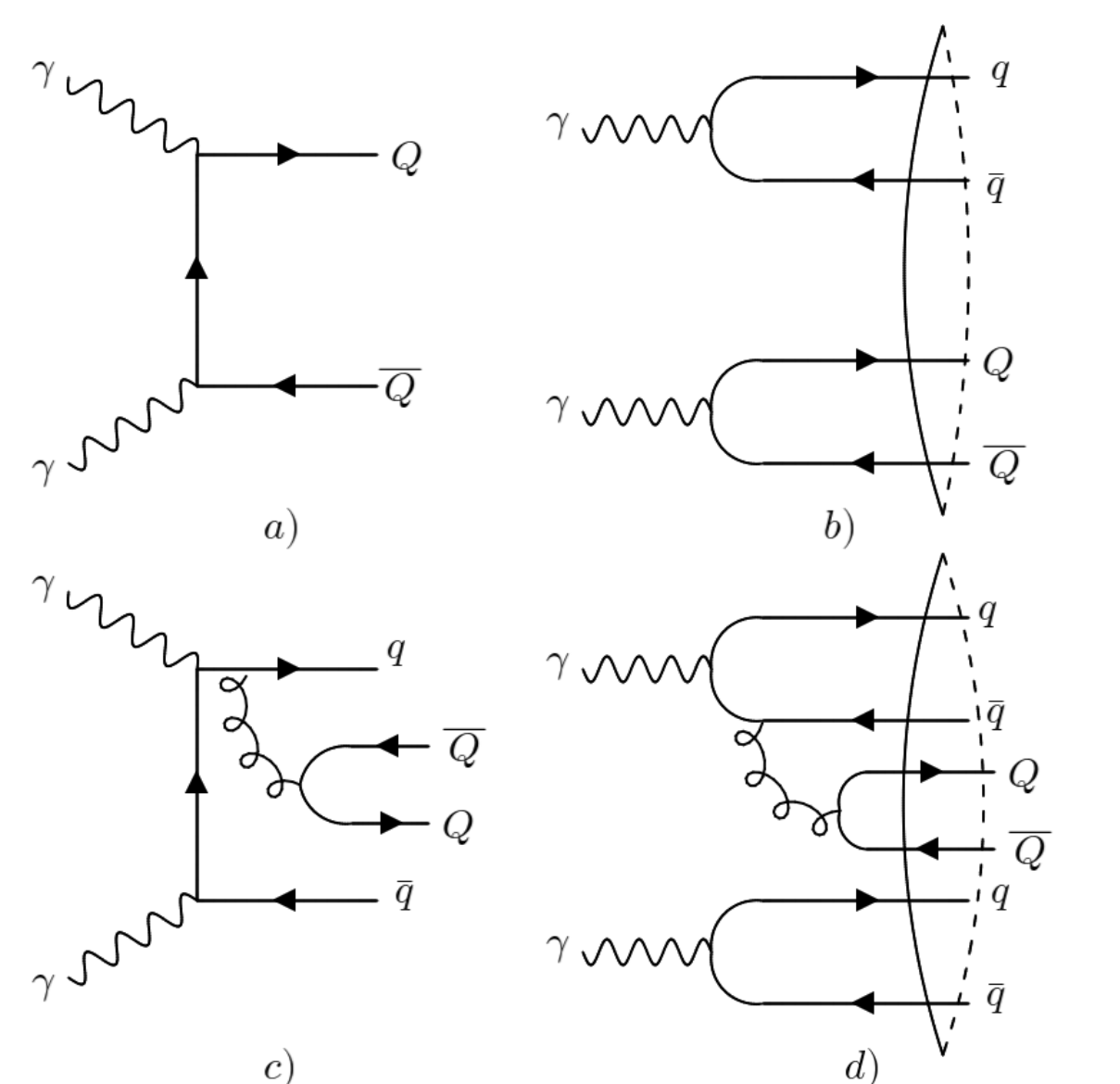}
\caption{Diagrammatic contributions to heavy-quark production.}
\label{fig:QPM_heavy_quarks}
\end{figure}

The diagrammatic contributions are shown in Fig.~\ref{fig:QPM_heavy_quarks}. Only diagrams (a) and (b) are included in our calculation; fragmentation diagrams (c) and (d) are omitted.

\subsection{Flavor contribution}
\begin{figure}[htbp]
\centering
\includegraphics[width=0.5\textwidth]{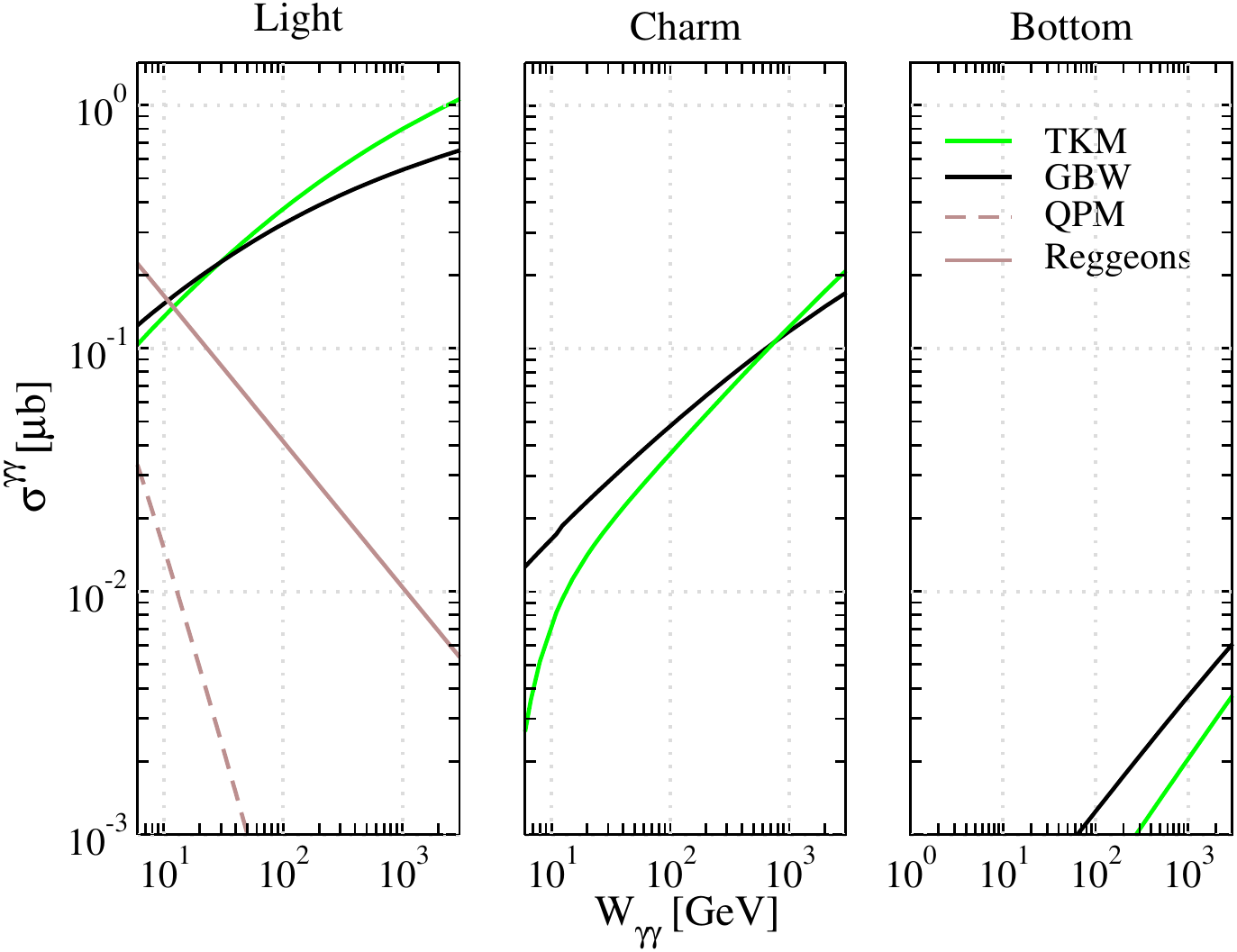}
\caption{Flavour contributions to $\sigma^{\gamma\gamma}$.}
\label{fig:flavor_contribution}
\end{figure}

Figure~\ref{fig:flavor_contribution} presents the flavour decomposition for the GBW model. The TKM prescription predicts larger light-quark production at intermediate energies but underestimates heavy-quark production relative to IKT. The Reggeon contribution remains comparable to bottom production even at high energies.

\subsection{Virtual photon cross section}
\begin{figure}[htbp]
\centering
\includegraphics[width=0.5\textwidth]{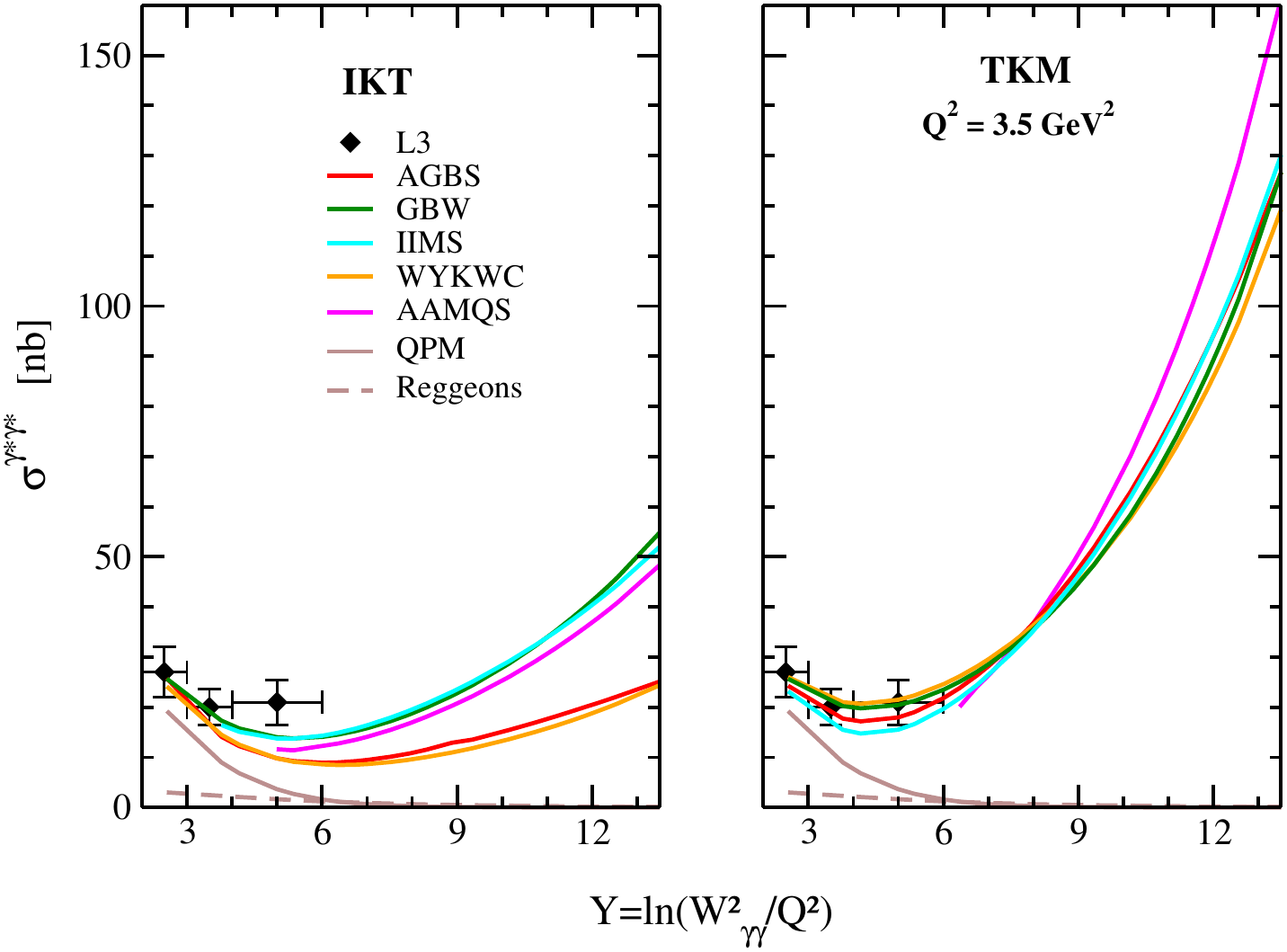}
\caption{$\sigma^{\gamma^*\gamma^*}$ at $Q^2=3.5$ GeV$^2$. Left: IKT; right: TKM.}
\label{fig:SV_3.5}
\end{figure}

\begin{figure}[htbp]
\centering
\includegraphics[width=0.5\textwidth]{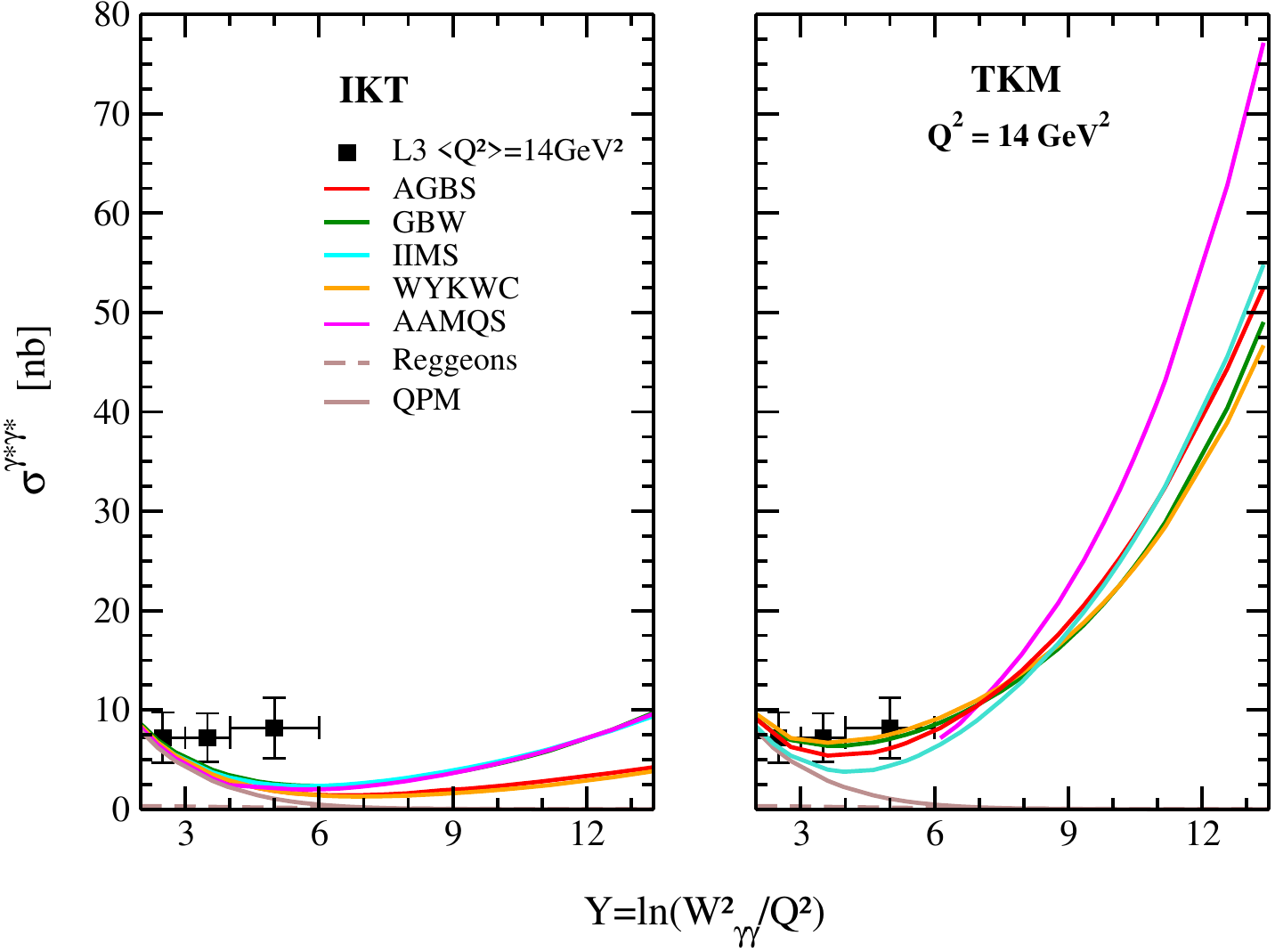}
\caption{$\sigma^{\gamma^*\gamma^*}$ at $Q^2=14$ GeV$^2$.}
\label{fig:SV_14.0}
\end{figure}

\begin{figure}[htbp]
\centering
\includegraphics[width=0.5\textwidth]{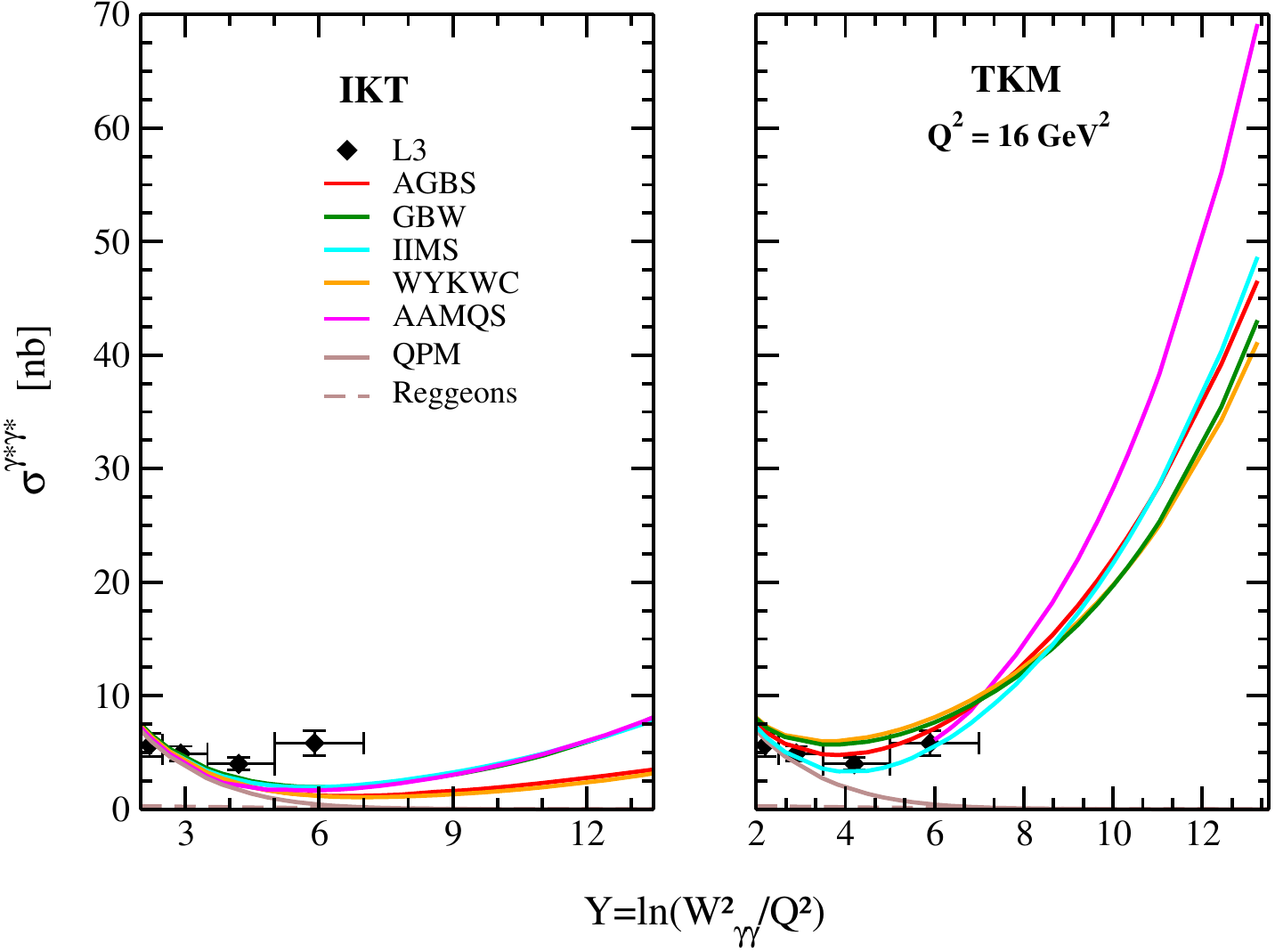}
\caption{$\sigma^{\gamma^*\gamma^*}$ at $Q^2=16$ GeV$^2$.}
\label{fig:SV_16.0}
\end{figure}

\begin{figure}[htbp]
\centering
\includegraphics[width=0.5\textwidth]{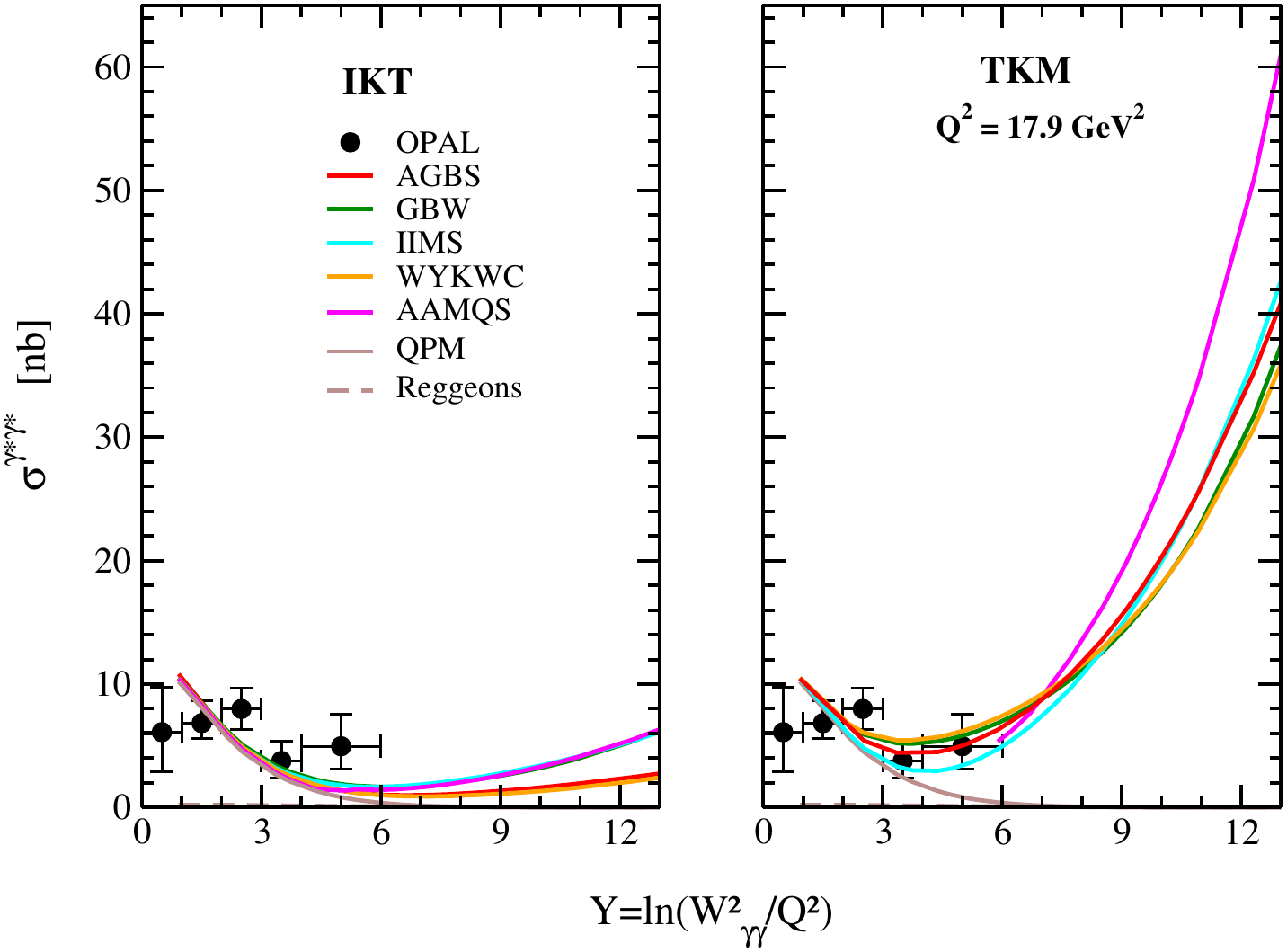}
\caption{$\sigma^{\gamma^*\gamma^*}$ at $Q^2=17.9$ GeV$^2$.}
\label{fig:SV_17.9}
\end{figure}

Figures~\ref{fig:SV_3.5} to \ref{fig:SV_17.9} show $\sigma^{\gamma^*\gamma^*}$ as a function of $Y=\ln(W^2/Q^2)$ for $Q^2=3.5,14,16,17.9$ GeV$^2$. The TKM prescription (right panels) gives larger values than IKT (left panels) at high $Y$, and the difference increases with $Q^2$. For $Q^2=17.9$ GeV$^2$, TKM predicts roughly twice the IKT cross section.


\subsection{Photon structure function $F_2^{\gamma}$}
\begin{figure}[htbp]
\centering
\includegraphics[width=0.5\textwidth]{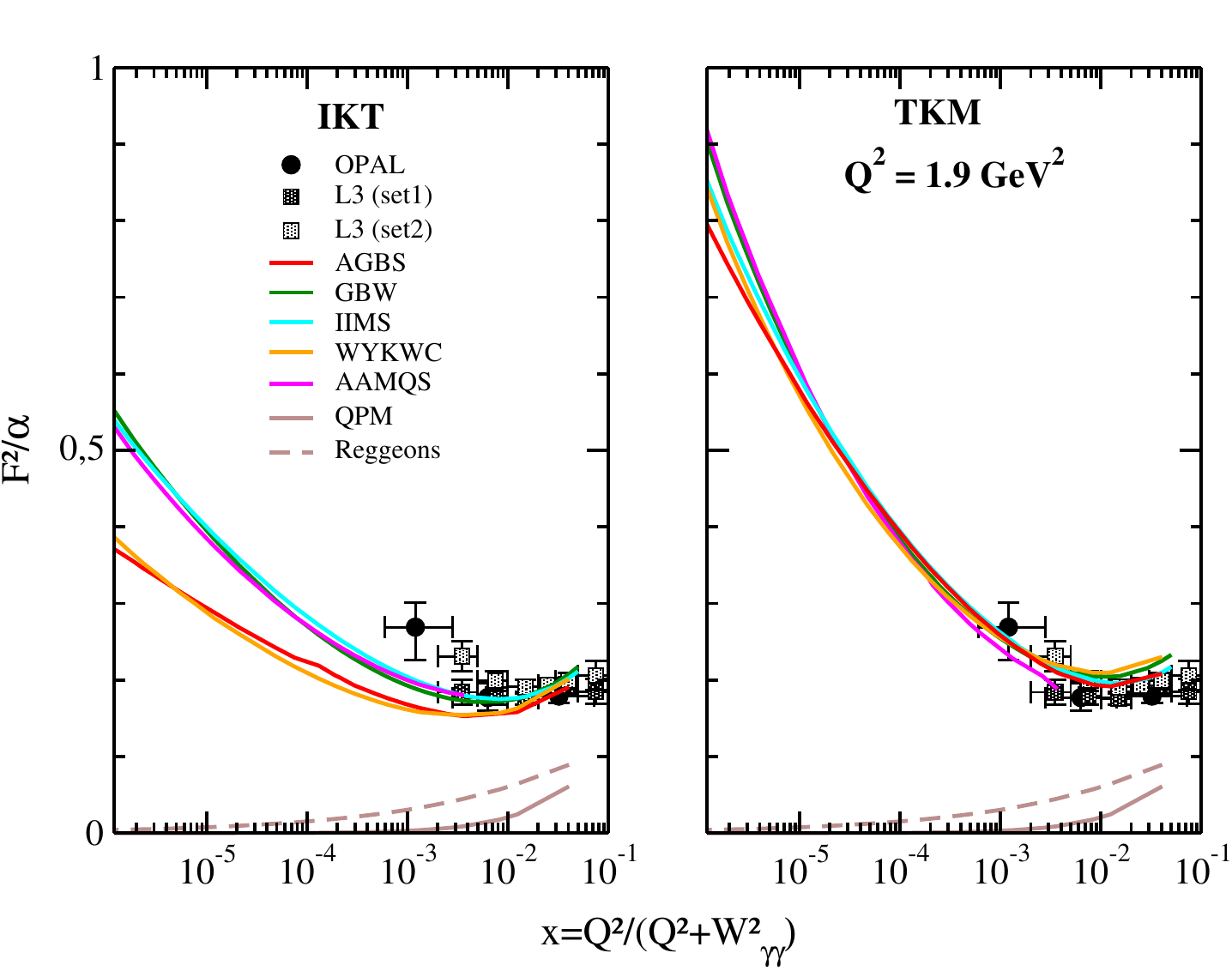}
\caption{$F_2^\gamma/\alpha$ at $Q^2=1.9$ GeV$^2$.}
\label{fig:F2_1.9}
\end{figure}

\begin{figure}[htbp]
\centering
\includegraphics[width=0.5\textwidth]{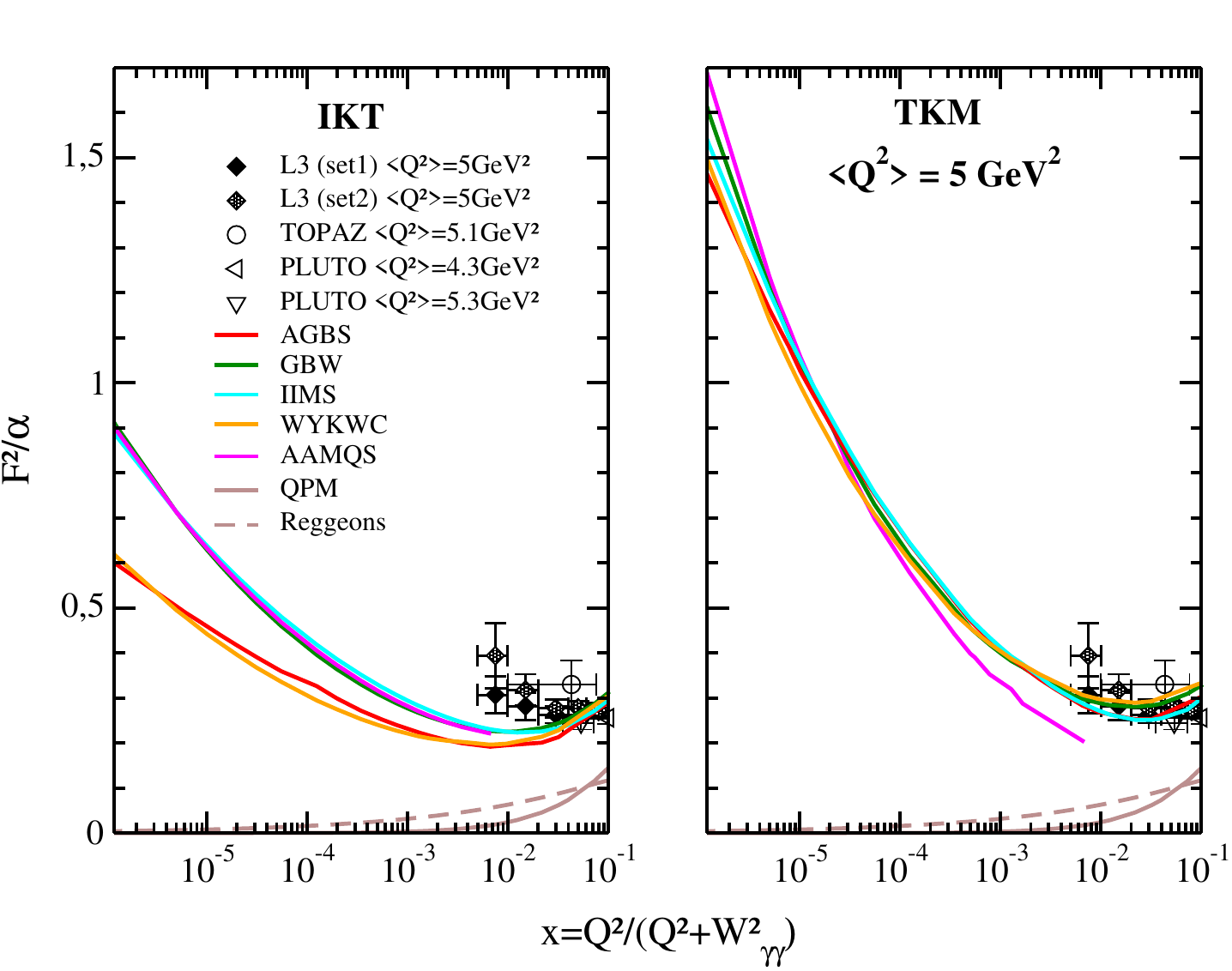}
\caption{$F_2^\gamma/\alpha$ at $Q^2=5.0$ GeV$^2$.}
\label{fig:F2_5.0}
\end{figure}

\begin{figure}[htbp]
\centering
\includegraphics[width=0.5\textwidth]{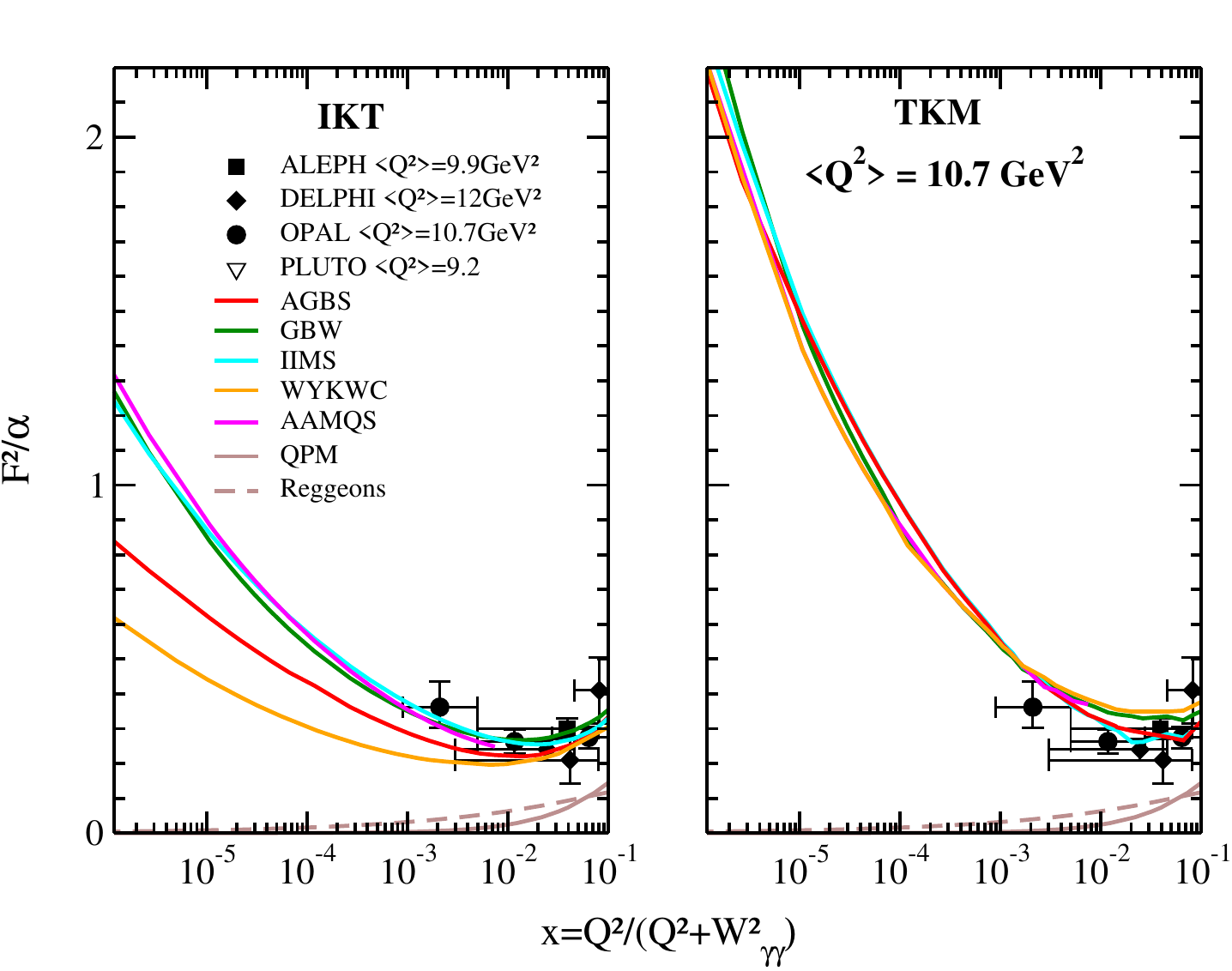}
\caption{$F_2^\gamma/\alpha$ at $Q^2=10.7$ GeV$^2$.}
\label{fig:F2_10.7}
\end{figure}

\begin{figure}[htbp]
\centering
\includegraphics[width=0.5\textwidth]{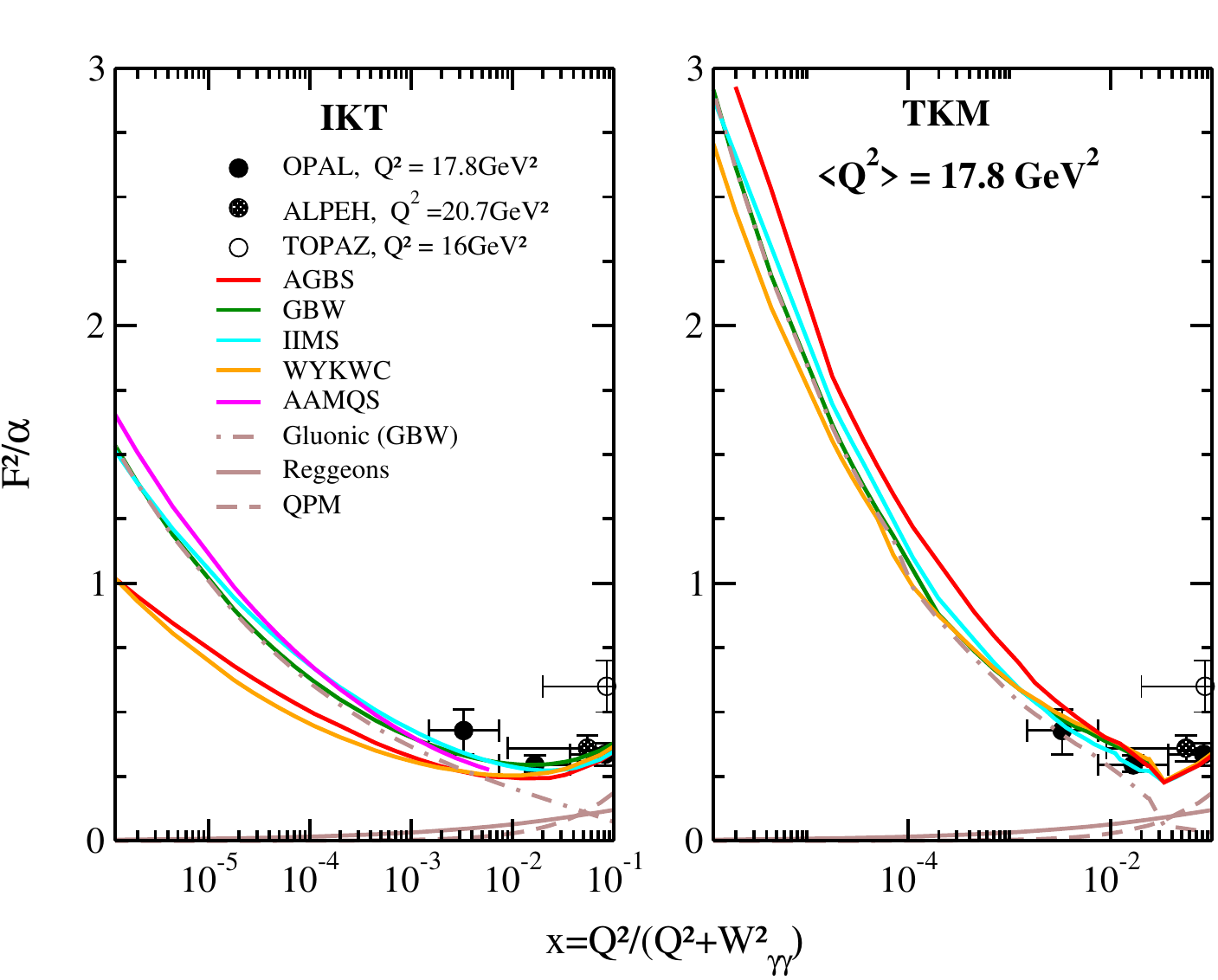}
\caption{$F_2^\gamma/\alpha$ at $Q^2=17.8$ GeV$^2$.}
\label{fig:F2_17.9}
\end{figure}

Figures~\ref{fig:F2_1.9} to \ref{fig:F2_17.9} show $F_2^\gamma/\alpha$ as a function of $x=Q^2/(W^2+Q^2)$. The TKM prescription (right) exhibits a steeper growth at small $x$ (higher energies) than IKT (left), reflecting a denser gluon distribution.


\subsection{Parton density and dipole size analysis}
\begin{figure}[htbp]
\centering
\includegraphics[width=0.43\textwidth]{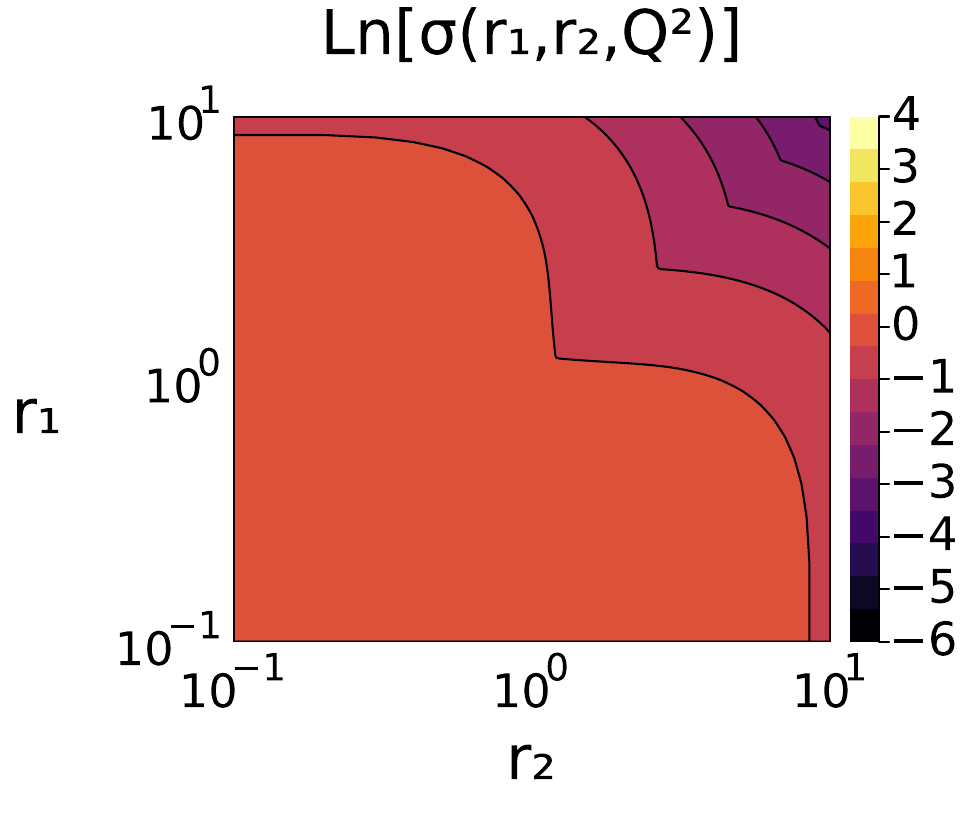}
\includegraphics[width=0.43\textwidth]{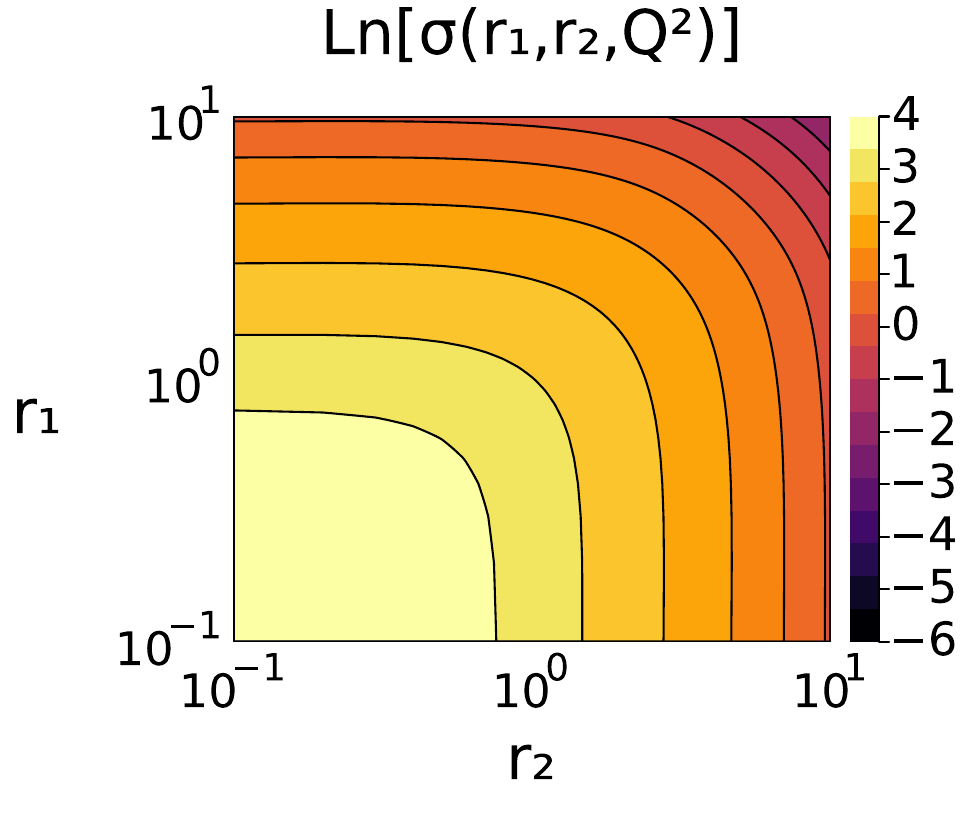}
\caption{Non-integrated gluonic cross section for real photons at $W=500$ GeV.}
\label{fig:NS_SR}
\end{figure}

\begin{figure*}[htbp]
\centering
\includegraphics[width=0.47\textwidth]{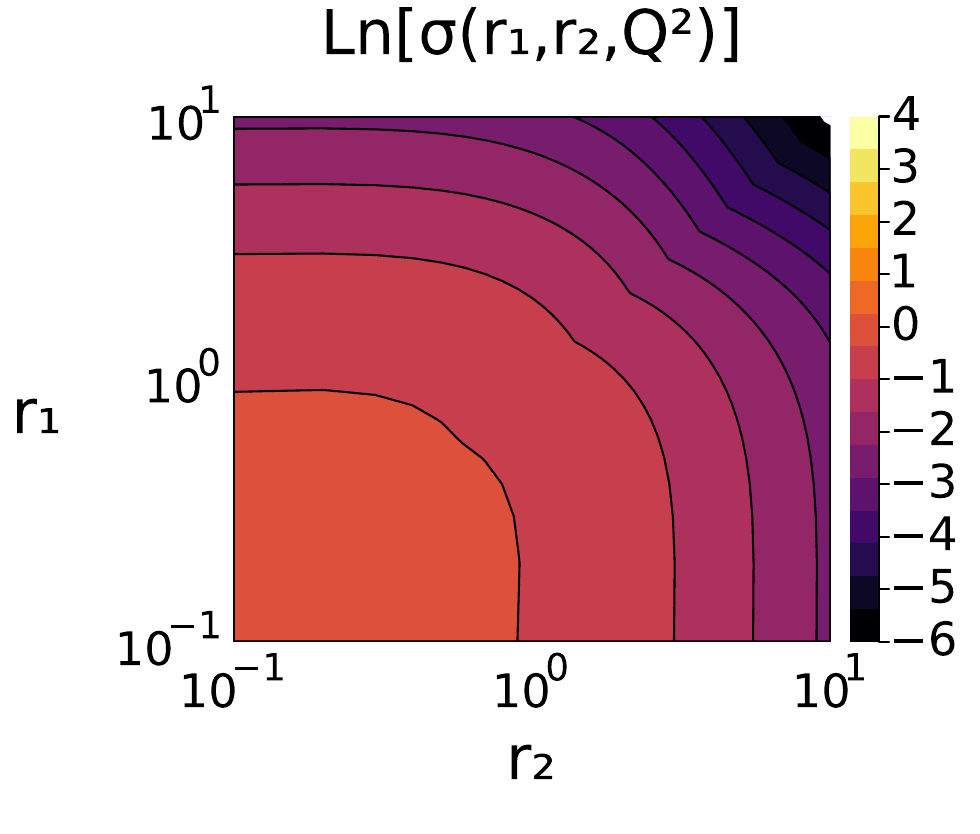}
\includegraphics[width=0.47\textwidth]{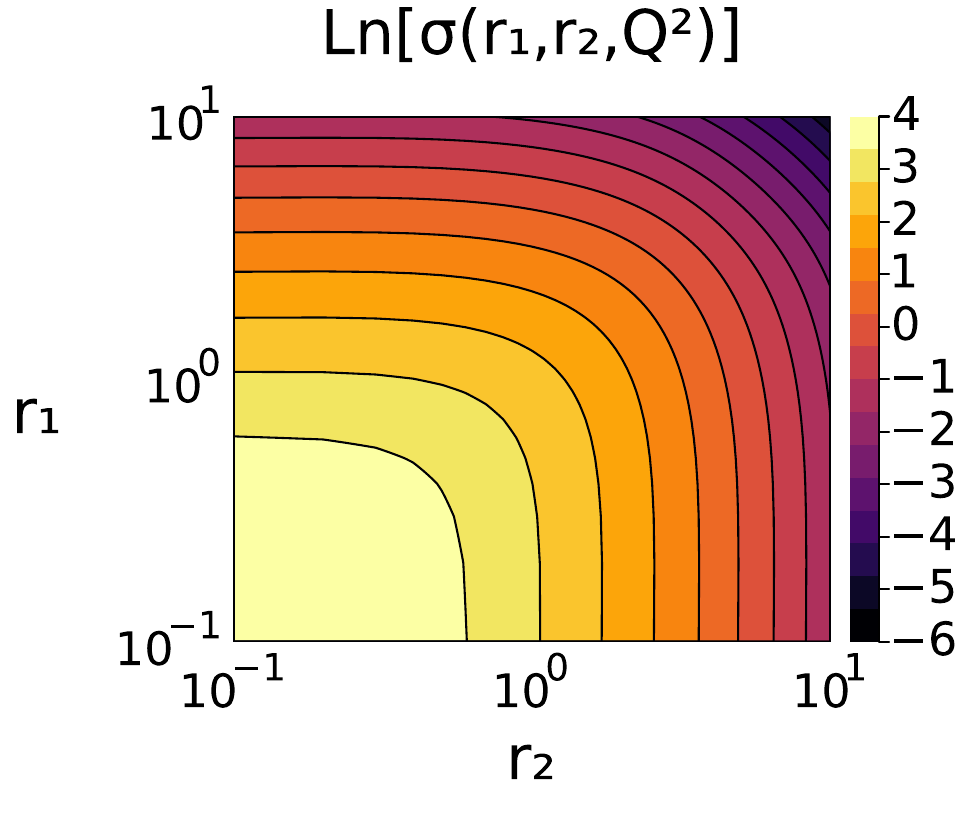}
\caption{Non-integrated virtual cross section at $Q^2=2$ GeV$^2$, $W=500$ GeV.}
\label{fig:NS_SV}
\end{figure*}

\begin{figure*}[htbp]
\centering
\includegraphics[width=0.47\textwidth]{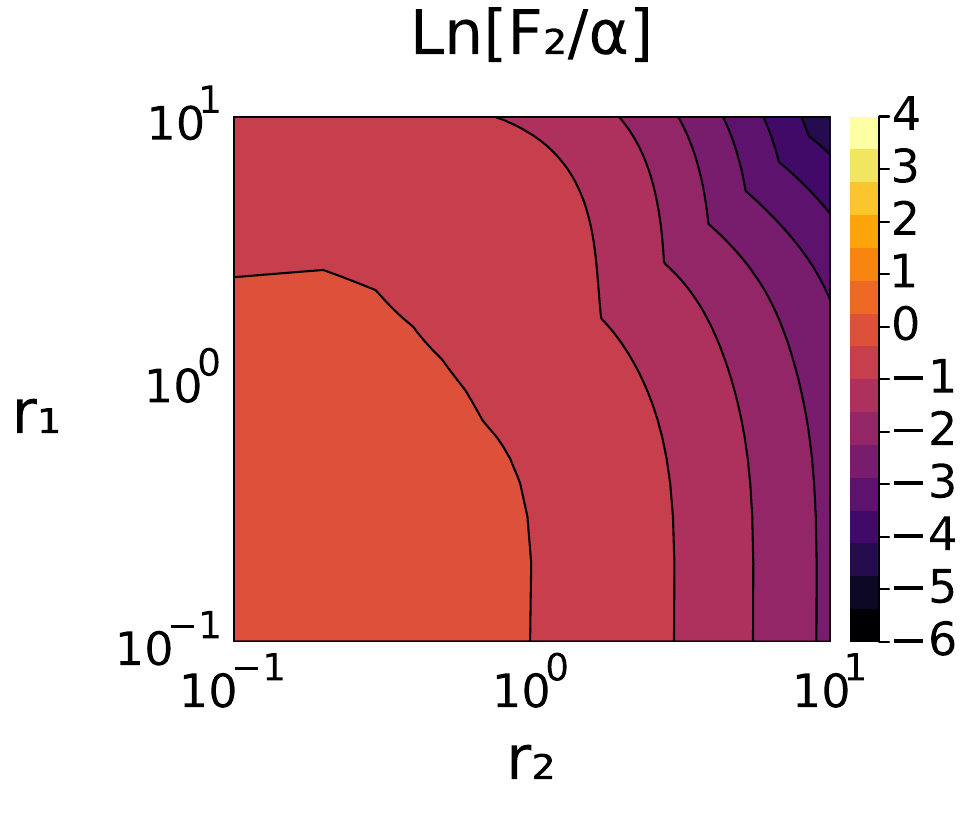}
\includegraphics[width=0.47\textwidth]{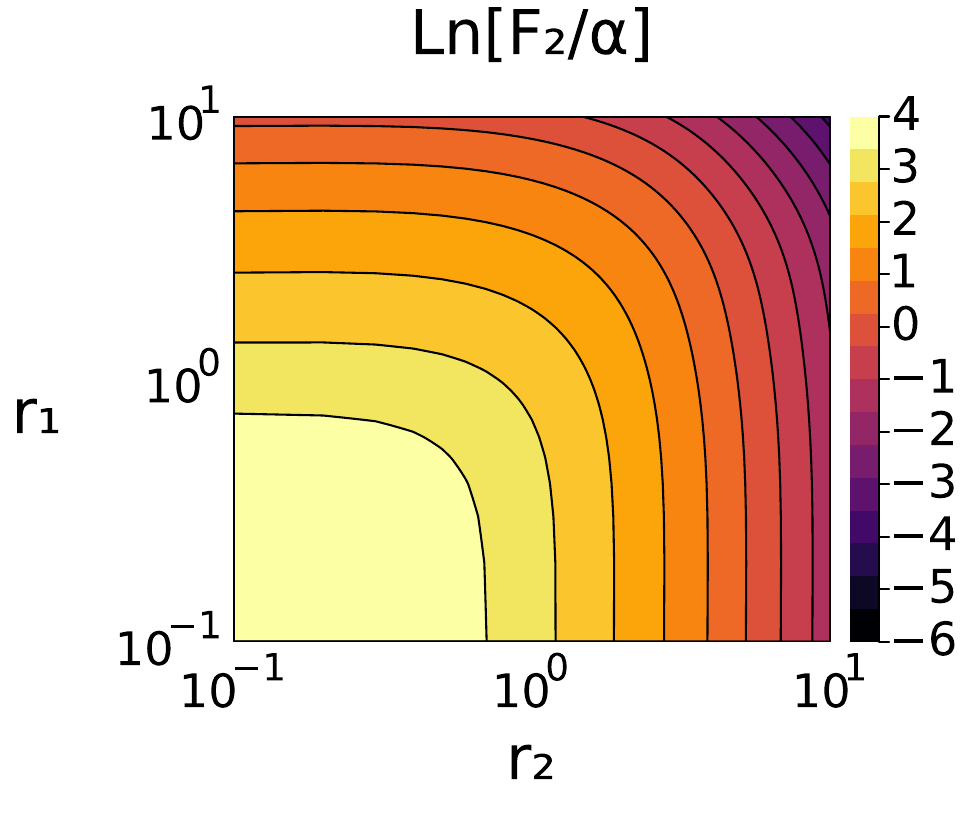}
\caption{Non-integrated photon structure function at $Q^2=2$ GeV$^2$, $W=500$ GeV.}
\label{fig:NS_F2}
\end{figure*}

To understand the origin of the TKM-IKT difference, we examine the non-integrated gluonic cross section $\sigma_G(r_1,r_2)$ in Fig.~\ref{fig:NS_SR} (real photons), Fig.~\ref{fig:NS_SV} (virtual photons), and Fig.~\ref{fig:NS_F2} (photon structure function). For TKM (right panels), the dominant contributions come from small dipoles ($r_{1,2}<1$ GeV$^{-1}$); for IKT (left panels), there are significant contributions from one large dipole as long as the other is small. This confirms that TKM describes a “dense+small” photon, while IKT describes a “dilute+large” photon.

\begin{figure}[htbp]
    \centering
    \includegraphics[width=1.0\linewidth]{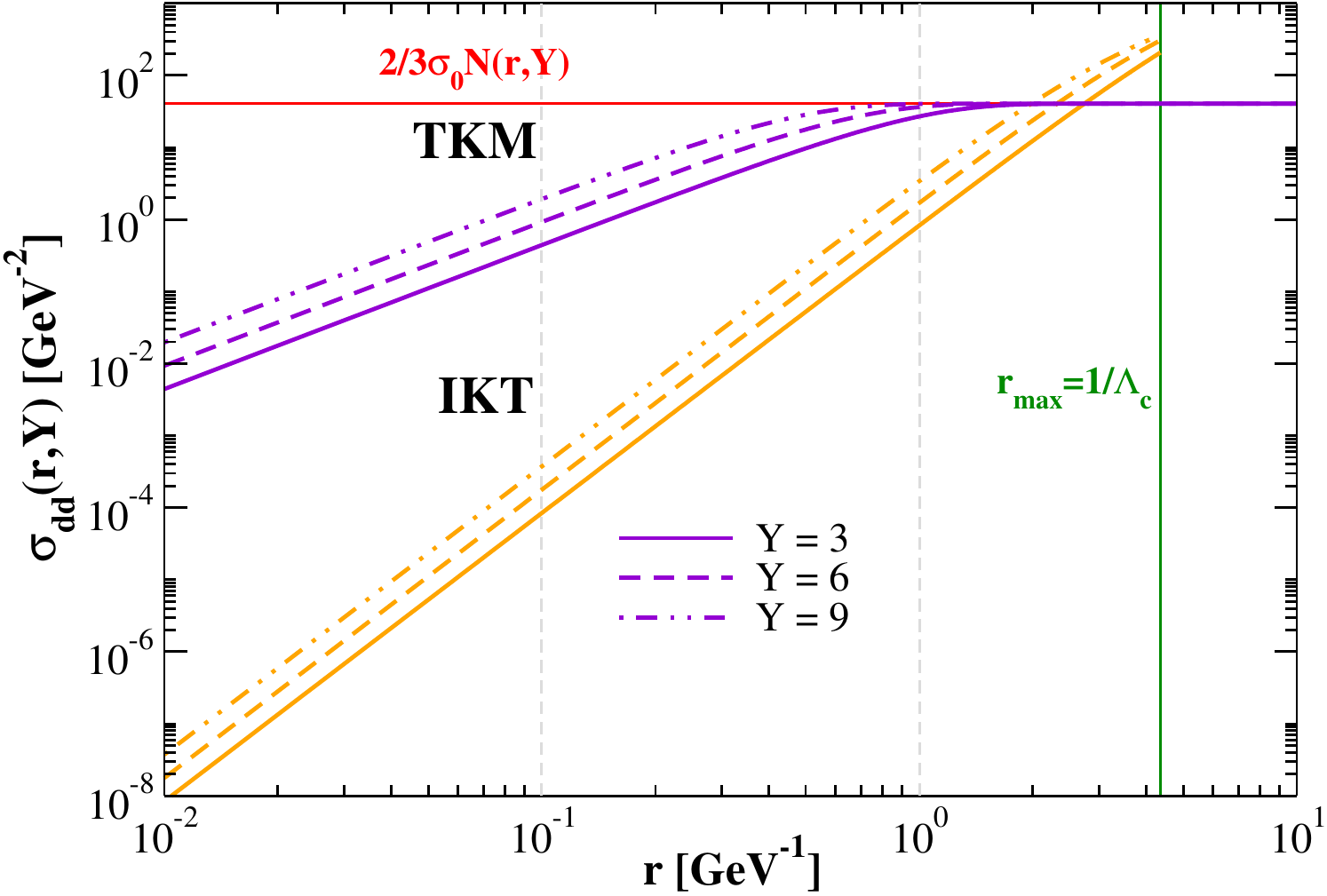}
    \caption{Dipole-dipole cross section vs. $r$ for different $Y$.}
    \label{fig:Sig_dd}
\end{figure}

Figure~\ref{fig:Sig_dd} compares the dipole-dipole cross section as a function of $r$ for TKM and IKT. TKM saturates to a constant at large $r$ (black-disk limit), while IKT does not reach the black-disk limit due to the cutoff $r_{\text{max}}$.

\subsection{Systematic uncertainties and model discrimination}\label{sec:syst}
The predictions presented above reveal a significant spread between the TKM and IKT prescriptions, especially at high energies and high virtualities.

To quantify this spread, Table~\ref{tab:delta} reports the relative difference \(\Delta\) averaged over the five dipole amplitudes, extracted from the benchmark dataset established in the companion paper \citep{ZardoBecker2026a}. The values at \(W = 100\) and \(200\) GeV are part of that benchmark.

\begin{table}[htbp]
\centering
\caption{Relative difference \(\Delta\) between TKM and IKT for real photons, extracted from the benchmark dataset established in the companion paper \citep{ZardoBecker2026a}.}
\begin{tabular}{c|c}
\hline
$W_{\gamma\gamma}$ (GeV) & $\Delta$ (\%) \\
\hline
100    & $42 \pm 8$ \\
200    & $48 \pm 12$ \\
\hline
\end{tabular}
\label{tab:delta}
\end{table}

Sensitivity studies reported in the literature \citep{tkm_2002, goncalves_amaral_2012} indicate that variations of \(\Lambda_c\) and \(m_l\) within typical ranges do not alter the qualitative hierarchy \(\sigma_{\text{TKM}} > \sigma_{\text{IKT}}\); hence the TKM–IKT spread is not an artifact of parameter tuning but a genuine structural ambiguity, already manifest in the benchmark dataset.

For virtual photons, Table~\ref{tab:delta_virt} shows \(\Delta\) at \(Q^2=2\) and \(10\)~GeV\(^2\), derived from the data at neighbouring virtualities (0.35,1.4,1.6,1.79~GeV\(^2\)) using geometric scaling (Eq.~(C.6)).

\begin{table}[htbp]
\centering
\caption{Relative difference \(\Delta\) for virtual photons, extracted from the benchmark dataset established in the companion paper \citep{ZardoBecker2026a}.}
\begin{tabular}{c|c|c}
\hline
$W_{\gamma\gamma}$ (GeV) & $Q^2=2$~GeV$^2$ (\%) & $Q^2=10$~GeV$^2$ (\%) \\
\hline
100    & $39.2$ & $36.5$ \\
200    & $44.8$ & $40.1$ \\
\hline
\end{tabular}
\label{tab:delta_virt}
\end{table}

To discriminate between the two pictures - TKM (dense+small dipoles) and IKT (dilute+large dipoles) - we propose three observables: (i) the energy growth of $\sigma^{\gamma\gamma}$, where TKM predicts a factor 2-3 larger cross section at $W=1$ TeV; (ii) the charm-to-bottom production ratio $R_{c/b} = \sigma^{c\bar{c}}/\sigma^{b\bar{b}}$, approximately $30\%$ higher in TKM; and (iii) exclusive vector meson pair production $\gamma\gamma \to J/\psi J/\psi$, which scales as $[\mathcal{N}(r,Y)]^2$ and thus amplifies the differences - TKM yields a cross section an order of magnitude larger than IKT at high $W$.

Future $e^+e^-$ colliders (ILC, CLIC, FCC-ee) with $\sqrt{s}\gtrsim 1$ TeV have the required luminosity to measure these quantities and settle the ambiguity.

\subsubsection{A suggestion for experimental comparison}
To facilitate a quantitative comparison between theory and experiment, we propose a dimensionless \textit{Discrimination Power (DP)}, defined as:
\[
DP_{O} = \frac{\left|O_{TKM} - O_{IKT}\right|}{\frac12\left|O_{TKM} + O_{IKT}\right|} \frac{1}{\delta_{exp}}
\]
where \(\delta_{exp}\) is the expected relative experimental uncertainty for a given observable \(O\) at a specific collider. The first factor is precisely the relative difference \(\Delta\) reported in Table~\ref{tab:delta}. A value \(DP \gtrsim 2\) implies that the model spread is at least twice the expected experimental error – sufficient for a clean discrimination. Experimental collaborations are encouraged to compute \(DP\) for the three proposed discriminants (\(\sigma^{\gamma\gamma}\), \(R_{c/b}\), \(\sigma^{\gamma\gamma\to J/\psi J/\psi}\)) using their anticipated luminosities and systematic error budgets, together with the benchmark values of \(\Delta\) reported in Table~\ref{tab:delta}.

\section{Conclusions}
\label{sec:conclusions}
This paper has translated the theoretical benchmark established in the companion work into a quantitative phenomenological baseline for high-energy two-photon interactions, evaluating real and virtual cross sections, the photon structure function, and inclusive heavy-quark production across five representative saturation models. By confronting the TKM and IKT dipole cross-section prescriptions, we have demonstrated that their systematic spread is by no means an artifact of parameter tuning, but rather an irreducible structural ambiguity. Specifically, for real photon total cross sections at $W_{\gamma\gamma} = 200\text{ GeV}$, this spread reaches $48\pm12\%$ and expands systematically into the multi-TeV domain. Our non-integrated gluon analysis confirms that this divergence reflects fundamentally distinct physical pictures: TKM enforces a ``dense+small'' photon configuration dominated by compact dipoles, whereas IKT accommodates a ``dilute+large'' photon structure.

Furthermore, analytical tests using DIS-tuned parameters for the GBW model indicate that the universality of the dipole amplitude is approximately satisfied—keeping predictions within a factor of 1.5–2 of LEP data without re-fitting—while leaving a residual discrepancy that demands future precision data. Inclusive heavy-quark production estimates via $K$-factors indicate that fragmentation may enhance direct cross sections by 20–50\% for charm and 20–30\% for bottom, serving as a conservative indicator of these channels' importance pending a full implementation of fragmentation functions within the dipole framework.

To resolve the TKM–IKT ambiguity beyond a mere choice of parameters, we have proposed three experimental discriminants for future $e^+e^-$ colliders with $\sqrt{s} \gtrsim 1\text{ TeV}$ (such as ILC, CLIC, and FCC-ee): (i) the energy growth rate of $\sigma^{\gamma\gamma}$, where TKM predicts a factor 2–3 larger value at $W = 1\text{ TeV}$; (ii) the charm-to-bottom production ratio $R_{c/b}$, which is approximately 30\% higher in TKM; and (iii) exclusive vector meson pair production $\gamma\gamma \to J/\psi J/\psi$. In particular, the latter scales quadratically as $[\mathcal{N}(r,Y)]^2$, magnifying the underlying divergence into an order-of-magnitude separation. Systematic numerical verification confirms that our results are robust against variations in integration tolerances and cutoffs (see Table~\ref{tab:convergence}), ensuring they are free from numerical artifacts.

\begin{table*}[htbp]
\centering
\caption{Numerical stability tests for the GBW model. The variations shown are indicative of the numerical precision of the benchmark calculations; they illustrate that the numerical uncertainties are small compared to the TKM–IKT spread reported in Table~\ref{tab:delta}.}
\begin{tabular}{lc}
\hline
Test & Variation \\
\hline
Tolerance \(10^{-4}\) (standard) & – \\
Tolerance \(10^{-5}\) & \(<0.1\%\) \\
\(r_{\min}=10^{-4}\ \mathrm{GeV}^{-1}\) & – \\
\(r_{\min}=10^{-3}\ \mathrm{GeV}^{-1}\) & \(<0.1\%\) \\
\(r_{\min}=10^{-2}\ \mathrm{GeV}^{-1}\) & \(<0.5\%\) \\
\(\Lambda_c\) nominal (IKT) & – \\
\(\Lambda_c -5\%\) & \(<2\%\) \\
\(\Lambda_c +5\%\) & \(<2\%\) \\
16 angular points & – \\
32 angular points & \(<0.1\%\) \\
\hline
\end{tabular}
\label{tab:convergence}
\end{table*}

Ultimately, these discriminants serve as rigorous falsification tests of the benchmark itself. Should upcoming data lie outside the TKM–IKT predictive envelope, it will signal the breaking point of classical premises—such as universal amplitude extraction, asymmetric rapidity evolution, and the homogeneous-disk approximation. These findings outline a clear theoretical roadmap, demonstrating that resolving these ambiguities will require moving beyond traditional heuristic prescriptions toward first-principles frameworks rooted in the JIMWLK functional.

\subsection{Open questions from this comparison}
The benchmark established in this work leaves at least three fundamental open questions directly tied to the proposed experimental discriminants:
\begin{enumerate}
    \item \textbf{Geometric scaling and photon saturation scale.} Validating whether the universal $Q_s(x)$ extracted from DIS governs $\gamma\gamma$ collisions tests the existence of a photon-specific saturation scale $Q_s^\gamma(x)$. Deviations observed through energy-dependent cross sections will directly falsify the strict universality hypothesis.
    \item \textbf{Symmetric vs. ordered dipole interactions.} The sensitivity of the $R_{c/b}$ ratio to dipole size composition tests whether a fully symmetric evolution ($Y_1 + Y_2$) replaces the need for asymmetric rapidity approximations.
    \item \textbf{Non-linear QCD dynamics in exclusive channels.} The quadratic sensitivity of $\gamma\gamma \to J/\psi J/\psi$ provides a powerful magnifying glass for high-density parton regimes, offering definitive constraints on non-linear evolution models.
\end{enumerate}
These open questions define a roadmap for future theoretical and experimental work. The discriminants proposed in this paper are designed to address them, and the benchmark established here provides the quantitative baseline against which future data can be compared.

\subsubsection*{Reproducibility}
To ensure reproducibility of the benchmark calculations, all model parametrizations, fitting procedures, and numerical integration settings are specified in the main text. The numerical code used for the calculations is available from the corresponding author upon reasonable request.

\subsubsection*{Acknowledgment}
The author would like to express special thanks to João Thiago de Santana Amaral for their extensive advice throughout the development of this work. Gratitude is extended to Yuri Kovchegov and Brandon Manley for their helpful discussions and valuable suggestions. This work was partially supported by the Brazilian funding agencies CNPq, CAPES, and INCT-FNA.

\nocite{*}
\bibliography{Zardo2AN}
\end{document}